\documentclass[preprint, times]{elsarticle}

\usepackage{amssymb,amsthm,amsmath,float}
\usepackage[total={6.5in,8.75in}, top=1.2in, left=0.9in, includefoot]{geometry}
\usepackage{graphicx}
\usepackage[pdftex,bookmarks,colorlinks,breaklinks]{hyperref} 

\newcommand{\Eq}[1]{(\ref{eq:#1})}

\newcommand{\Sec}[1]{\S \ref{sec:#1}}
\newcommand{\Fig}[1]{Fig.~\ref{fig:#1}}

\newcommand{\InsertFig}[4]
{\begin{figure}[ht!]		
       \centerline{
         \includegraphics[width=#4]{#1}
       }
       \caption{{\footnotesize  #2}
       \label{fig:#3}}
\end{figure}}

\newcommand{\bR}{{\mathbb{ R}}}

\newcommand{\bZ}{{\mathbb{ Z}}}

\newcommand{\cO}{{\cal O}}
\newcommand{\co}{{o}}

\newcommand{\cS}{{\cal S}}

\newcommand{\sgn}{\mathop{\rm sgn}\nolimits}

\newcommand{\adj}{\mathop{\mathrm{adj}}\nolimits}

\newtheorem{thm}{Theorem}

\newcommand{\beq}[1]{\begin{equation}\label{eq:#1}}
\newcommand{\eeq}{\end{equation}}

\newcommand{\bsplit}[1]{\begin{equation}\label{eq:#1}\begin{split}}
\newcommand{\esplit}{\end{split}\eeq}

\def\bcb{border-collision bifurcation}
\def\cS{\mathcal{S}}

\def\db{discontinuous bifurcation}

\def\pwl{piecewise-linear}
\def\pws{piecewise-smooth}

\def\sew{symbol sequence}
\def\shr{shrinking point}
\def\sL{{\sf L}}
\def\sR{{\sf R}}
\def\sw{switching manifold}
\def\tong{resonance tongue}

\begin{document}
\begin{frontmatter}
\title{Aspects of Bifurcation Theory for\\ Piecewise-Smooth, Continuous Systems}
\author[ubc]{D.J.W.~Simpson\fnref{fn1}}
\ead{david.jw.simpson@gmail.com}

\author[cub]{J.D.~Meiss\corref{cor}\fnref{fn2}}
\ead{James.Meiss@colorado.edu}

\cortext[cor]{Corresponding author}
\fntext[fn1]{DJWS acknowledges support from an NSERC Discovery Grant.}
\fntext[fn2]{JDM acknowledges support from NSF grant DMS-0707659.}

\address[ubc]{Department of Mathematics, University of British Columbia, Vancouver, BC, V6T1Z2, Canada}
\address[cub]{Department of Applied Mathematics, University of Colorado, Boulder, CO, 80309-0526, USA}

\date{\today}

\begin{abstract}
Systems that are not smooth can undergo bifurcations that are forbidden in smooth systems. We review some of the phenomena that can occur for piecewise-smooth, continuous maps and flows when a fixed point or an equilibrium collides with a surface on which the system is not smooth. Much of our understanding of these cases relies on a reduction to piecewise linearity near the border-collision. We also review a number of codimension-two bifurcations in which nonlinearity is important.  
\end{abstract}


\end{frontmatter}

\section{Introduction}

A dynamical system is {\em piecewise smooth} (PWS) if its phase space can be partitioned into countably many regions within which it is smooth. The codimension-one sets on which the dynamics is not smooth are called {\em switching manifolds}. The recent explosion of interest in these systems is demonstrated by the rapid growth of publications in the past two decades shown in \Fig{ISINonsmooth}, the appearance of a number of texts 
\cite{Br99, BaVe01, ZhMo03, Ts03, LeNi04, DiBu08, Si10}, as well as this issue of Physica D.

\InsertFig{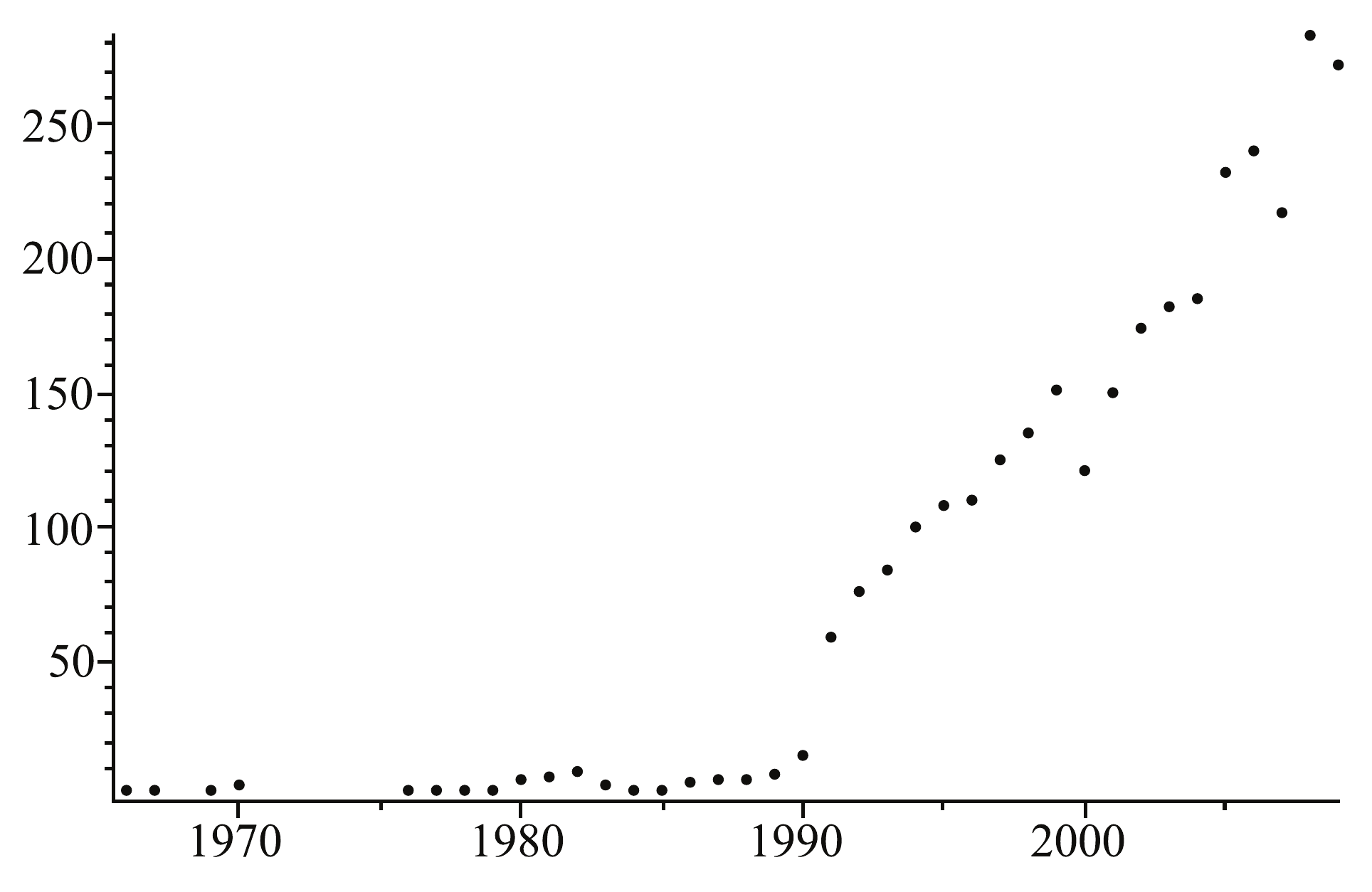}{Number of papers published each year on piecewise smooth dynamics from a search of the ISI Web of Knowledge\textregistered~database. Keywords were Topic=(piecewise smooth) OR (discontinuity induced) OR (nonsmooth) OR (grazing bifurcation) OR (border-collision bifurcation) AND Topic=(dynamics).}{ISINonsmooth}{3in}

Classical bifurcation theory relies heavily on the assumption of a certain degree of smoothness in the singularity conditions (e.g. zero eigenvalues), the nondegeneracy requirements (nonvanishing coefficients of some terms in a power series), and the use of the center manifold theorem (to achieve a dimension reduction).  Bifurcations can occur in nonsmooth systems that are forbidden in smooth systems; those that occur due to the interaction of invariant sets with a switching manifold are known as {\em discontinuity induced bifurcations}.
A simple example corresponds to the PWS continuous, one-dimensional family of ``tent maps", 
\beq{tent}
	T(x ; r) = \left\{\begin{array}{ll}
		rx, & x< \frac12 \\
		r(1-x), & x \ge \frac12
		\end{array}\right. \;,
\eeq
for which infinitely many periodic orbits are created as $r$ increases through $1$, without the usual sequence of period doubling bifurcations of smooth maps
\cite{Ra90,Gl99}.\footnote
    {The general \pwl, one-dimensional map was treated by 
    Nusse and collaborators \cite{NuYo95} in their pioneering papers on 
    border-collision bifurcations, see also \cite{DiBu08}.}

Of the many types of discontinuities that can occur, perhaps the simplest is a discontinuity in the first derivative; in this case the dynamical system is piecewise-smooth and continuous (PWSC). The simplest discontinuity induced bifurcations in these systems correspond to an equilibrium (for ODEs) or a fixed point (for maps) that lies on the switching manifold. The resulting bifurcations are known as {\em discontinuous} or {\em border-collision} bifurcations, respectively. The bifurcations occurring in  this class of systems are reviewed in this paper.

More extreme discontinuities include {\em hybrid systems} in which the dynamics may be defined by different types of models in different regions \cite{VaSc00, Jo03}, and {\em Filippov systems} where the vector field is discontinuous on a switching manifold and the behavior may include sliding along the manifold \cite{Fi88}.

In this paper we will review some recent results on low-codimension discontinuous and border-collision bifurcations when the switching manifold is smooth at the bifurcation point.\footnote
	{Bifurcations at corners of a \sw~are discussed in \cite{LeNi04, LeVa06}.}
One of the difficulties in treating these bifurcations generally corresponds to the absence of a center manifold reduction. Even though there are still many open questions, there are a number of cases of $N$-dimensional flows and maps for which general bifurcation results have been obtained.

We focus on PWSC systems whose components are $C^k$
(for some $k \in \mathbb{N}$). With this restriction we disregard, for instance, maps with  square-root terms that arise from regular grazing in Filippov systems \cite{DiBu01,DiBu08}; however, we are able to exploit a major simplification:
\bcb s can often be treated with a \pwl~model.
For example, nonlinear terms do not influence generic discontinuous
saddle-node-like and Hopf-like bifurcations.
An important consequence is that the size of invariant sets
created in these bifurcations grows linearly with parameters
instead of as some fractional power for the smooth case.

Piecewise-smooth dynamics arises in many areas including electronics \cite{Ts03}, control theory \cite{AlBe01}, impacting mechanical systems \cite{LeNi04,Po00}, economics \cite{PuSu06,CaJa06,LaMo06} and biology \cite{Ro70,CaDe06,ZhSc08}---an extensive set of references can be found in \cite{DiBu08, Si10}. While continuous models are not always appropriate for these applications, there are many cases for which the model is given by either a continuous vector field or map. For example, PWSC dynamics occurs in the well-known Chua circuit, introduced in 1983, that contains a diode often modeled with a piecewise-linear, current-voltage response curve \cite{Ch94}. Similarly power circuitry like DC/DC buck/boost converters, used to decrease or increase voltage levels in common devices such as cell phones, can often be modeled by PWSC maps \cite{YuBa98}. Piecewise smooth, continuous economic models include discrete-time models for trade \cite{SuGa06}, production-distribution \cite{MoLa07} and of a socialist economy \cite{HoNu95}. Mechanical systems with impacts, or with transitions from static to sliding friction provide another large application area. Continuous models often arise in these systems when a body grazes a rigid surface, leading to so-called grazing bifurcations \cite{DiBu01, ChOt94}. Biological models that are PWSC include models of neurons \cite{ChJu04}, neural networks \cite{CoOs00}, and cybernetic models which reduce complex chemical models by selecting optimal reaction pathways \cite{SiKo09}.

In the next section we introduce equations ideal for analyzing
local bifurcations of PWSC systems.
We solve for the equilibria or fixed points for these equations and describe
the observer canonical form. Section \ref{sec:continuousTime} summarizes recent results regarding codimension-one and two bifurcations for PWSC vector fields.
A discussion for the discrete-time scenario
is presented in \Sec{discreteTime}.

\section{Local Framework}\label{sec:frameWork}

Near a sufficiently smooth, codimension-one \sw~one
can always choose coordinates
so that the manifold is represented by the vanishing of one of
the coordinates, say the first coordinate \cite{DiBu01b}.
To simplify the notation, for any point $x \in \bR^N$ we let
\beq{s}
s = e_1^{\sf T} x 
\eeq
denote the first element of $x$; consequently, the \sw~is the plane $s=0$.
In this case a PWSC dynamical system takes different forms
in the left-half space, $s < 0$,
and the right-half space, $s > 0$:
\beq{pws}
	f(x;\xi) = \left\{ \begin{array}{ll}
					f^{(\sL)}(x;\xi), & s \le 0 \\
					f^{(\sR)}(x;\xi), & s \ge 0 
			  \end{array} \right. .
\eeq
Here the functions $f^{(i)}: \bR^N \times \bR^M \to \bR^M$, $i \in \{\sL,\sR\}$ are $C^k$ functions on the $N$-dimensional phase space with coordinates $x$ and parameters $\xi \in \bR^M$. We will assume that $k \ge 1$, though often the assumed $k$ will need to be larger.
The function $f$ may represent a vector field, in which case the dynamics is the system of ODEs
\beq{ode}
	\dot x = f(x;\xi) \;,
\eeq
or a map, in which case the dynamics is the discrete-time system
\beq{map}
	x \mapsto x' = f(x;\xi)\;.
\eeq
An orbit in the discrete case is a sequence $\{x_t: x_{t+1} = f(x_t,\xi), t\in\bZ\}$.
We will discuss these separately in \Sec{continuousTime} and \Sec{discreteTime}, respectively.

For the simplest bifurcations, we assume that the system has an equilibrium or fixed point on the switching manifold at some specific parameter value. Without loss of generality this point can be taken to be the origin, and we can write the parameters as $\xi = (\mu, \eta)$ where $\mu \in \bR$ represents a parameter that vanishes at the border collision. Then we can write
\beq{linearDef}
	f^{(i)}(x;\mu,\eta) = \mu b(\mu,\eta) + A_i(\mu,\eta) x + \cO(|x|^2) + \co(k) \;,
\eeq
where $A_\sL$ and $A_\sR$ are the $N \times N$ Jacobian matrices at the origin. Under the assumption of continuity the two vector fields must agree whenever $s = 0$, which implies that all of the columns of these two matrices must be identical except for the first
\[
	A_\sL e_i = A_\sR e_i \;,\quad i \neq 1 \;.
\]

The parameter $\mu$ corresponds to the primary parameter that unfolds a simple discontinuous or border-collision bifurcation, while $\eta$ will be used to represent parameters unfolding higher codimension cases. For simplicity, we suppress the dependence of the system on $\eta$, though the discussion will refer to these additional parameters as needed.

The nonlinear terms in \Eq{linearDef} can be regarded as higher-order in the sense that any structurally stable dynamics of the \pwl~system will persist when the nonlinear terms are added. When $\mu \ll 1$, it is useful to consider a scaled system, defining $z = \frac{x}{|\mu|}$ when $\mu \ne 0$ and $z = x$ when $\mu = 0$ to obtain
\beq{pwl}
	\dot{z} = \left\{ \begin{array}{ll}
					\sigma b(0) + A_\sL(0) z, & s < 0 \\
					\sigma b(0) + A_\sR(0) z, & s > 0
			  \end{array} \right. ,
\eeq
where $\sigma = \sgn(\mu)$ (and is $0$ when $\mu =0$). A nice feature of \Eq{pwl} is that $\mu$ has been replaced by the discrete parameter $\sigma \in \{ -1,0,1\}$.
Note that any structurally stable, bounded invariant set of \Eq{pwl} will approximate an invariant set of \Eq{pws} as $\mu$ tends to zero, and moreover that its size (in $x$) will shrink linearly to zero. This simple feature is one of the most prominent characteristics of discontinuous and border-collision bifurcations. 

If the matrices $A_i(0)$ are nonsingular, the implicit function theorem implies that the ODE \Eq{ode} has two potential equilibria, 
\beq{equil}
	x^{*(i)}(\mu) 
				 = -\frac{\varrho^{\sf T}(0) b(0)}{\det(A_i(0))} \mu + \cO(\mu^2)\;,
\eeq
where
\beq{rhodefine}
	 \varrho^{\sf T} = e_1^{\sf T} \adj(A_\sL(\mu)) = e_1^{\sf T} \adj(A_\sR(\mu)))\;,
\eeq
and $\adj(A)$ is the adjugate of $A$ (defined by $A \adj(A) = \det(A) I$ and extended by continuity for $\det(A) = 0$).
Note that \Eq{equil} also pertains to the map \Eq{map} if we replace $A_i$ by $I-A_i$ \cite{Si10}.

The equilibria \Eq{equil} are said to be {\em admissible}
when $s^{*(\sL)} \le 0$ and $s^{*(\sR)} \ge 0$.
Otherwise they are {\em virtual}.
Even virtual equilibria play a prominent role in the dynamics of \Eq{pwl},
since the eigenspaces of such equilibria typically
intersect the admissible half-space and these
intersections are therefore (forward or backward)
invariant sets of the full system.

A simple consequence of \Eq{equil} was first noted by Mark Feigin \cite{Fe74} (for the map case): there are two distinct types of border collision in this nonsingular case: if $\det(A_\sL(0))$ and $\det(A_\sR(0))$  (or $\det(I-A_\sL(0))$ and $\det(I-A_\sL(0))$ for the map case) have the same sign, then the equilibria \Eq{equil} are admissible for different signs of $\mu$, so that near the origin the system \Eq{pws} has a unique equilibrium. On the other hand if the signs of the Jacobian determinants differ, then the equilibria are admissible for only one sign of $\mu$ and they collide and annihilate as $\mu \to 0$, giving rise to a {\em nonsmooth fold} bifurcation.

Additional coordinate transformations can be performed on \Eq{pwl} to reduce the number of parameters.  Normally one could hope to transform the matrices to Jordan form; however, this cannot always be done simultaneously for both matrices, nor would it typically maintain the definition of the switching manifold, $s = 0$. However, under the condition of {\em observability}, the matrices can be reduced to {\em companion form}
\[
		C =  \left[\begin{array}{c|c}
			\begin{array}{c} -\Delta  \end{array}
			&
			\begin{array}{c} I \\ 0 \end{array}
		\end{array}\right] \;,
\]
where $\Delta$ is a vector and $I$ is the $(N-1) \times (N-1)$ identity matrix. Note that the components of $\Delta$ are the coefficients of the characteristic polynomial $p(\lambda) = \det(\lambda I -C)$.
Consequently every matrix has a unique companion form.
\begin{thm}[Observer Canonical Form \cite{DiBu08,Si10}]
There is a coordinate transformation that reduces
\Eq{pws} with \Eq{linearDef} to the {\em observer canonical form}:
\Eq{pws} with
\beq{companion}
	\hat{f}^{(i)}(x;\mu) = \mu e_N + C_i(\mu) x + \cO(|x|^2) + \co(k) \;,
\eeq
for small $\mu$ and where $C_i$ is the companion matrix of $A_i$,
if and only if $A_\sL(0)$ has no eigenspace orthogonal to $e_1$.
\end{thm}

Note that the switching manifold for the observer form is still $s = 0$. The \pwl~version of \Eq{companion} has $N$ parameters for each matrix, so its complete bifurcation picture is determined by $2N$ parameters, in addition to $\mu$.

\section{Continuous-Time Systems}\label{sec:continuousTime}

The determination of the stability of an equilibrium located on
a \sw~is an important and fundamental yet difficult problem
in nonsmooth systems \cite{CaDe06,LiAn09}.
For PWSC systems, stability depends on
the eigenspaces of each adjoining smooth subsystem
and their relative configuration with respect to \sw(s),
and, in special cases, nonlinear terms of the system \cite{GoMe03, IwHa06}.
In the absence of nonlinear terms,
radial symmetry allows for a projection of the dynamics onto a space of dimension
one less than the dimension of the system,
which is a useful simplification \cite{Ha92}.

When there is a single \sw, as for \Eq{pws}, in one or two dimensions
the stability problem is straightforward \cite{FrPo98}.
In three or more dimensions the problem is significantly more complicated.
For example, an equilibrium on a \sw~may
be unstable even if all eigenvalues of
both $A_\sL$ and $A_\sR$ have negative real part \cite{CaFr06}.
There may exist cone-shaped invariant sets
akin to slow eigenspaces of smooth systems \cite{CaFr05,Ku08}. By contrast,
in two dimensions such an equilibrium undergoes a Hopf-like bifurcation
when, roughly speaking, trajectories of the linearization
spiral inwards as much as they spiral
outwards, per rotation \cite{KuMo01,ZoKu05,ZoKu06}.
These considerations are important for some models of real-world systems for which some natural constraint results in an equilibrium that remains on a
\sw~under parameter perturbation \cite{Ku08,AsTa09th}.

Bifurcations of equilibria in smooth systems of arbitrary dimension
are readily classifiable as saddle-node, Hopf, and so forth,
because the degenerate dynamics occur on a low dimensional subspace (center manifold).
For the PWSC system, \Eq{pws}, this dimension reduction 
cannot be performed because
the system is not differentiable in a neighborhood of the equilibrium.
For this reason \db s resist a simple classification.
Though discontinuous bifurcations may generate complex dynamics,
such as Silnikov homoclinic chaos \cite{Sp81,LlPo07},
high-dimensional, PWSC systems often can exhibit only low-dimensional dynamics. It remains a difficult problem to develop a general theory of dimension reduction
for such systems.

One idea is to look at the eigenvalues of the convex hull of the
two linearizations of the equilibrium \cite{LeNi04,Le06}:
\[
	\hat{A} = \left\{ (1-\sigma) A_\sL(0) + \sigma A_\sR(0) ~\Big|~ \sigma\in [0,1] \right\} \;,
\]
For $\sigma \in [0,1]$,
the eigenvalues of $\hat{A}$ provide a continuous connection
between the eigenvalues of $A_\sL(0)$ and $A_\sR(0)$
and are collectively referred to as the {\em eigenvalue path}.
The idea is that a bifurcation occurs if and only if
the eigenvalue path intersects the imaginary axis,
though currently this remains a conjecture.
Furthermore, examples indicate that the complexity of the
\db~increases with the number of intersections between the eigenvalue path
and the imaginary axis.
In particular, the presence of exactly one intersection
typically yields a \db~that is a natural
analogue of a familiar smooth bifurcation.
Nonsmooth counterparts of saddle-node, Hopf, pitchfork and transcritical
bifurcations have been studied \cite{LeNi04}.
The main difference is that invariant sets created at nondegenerate
\db s grow in size linearly, to lowest order.

When \Eq{pws} is planar, all possible codimension-one, \db s
that may occur have been determined \cite{LuCh91,FrPo98}.
If the equilibria $x^{*(\sL)}$ and $x^{*(\sR)}$, \Eq{equil}, are nondegenerate they
may be classified as saddles, attracting nodes, attracting foci, repelling foci or repelling nodes. The bifurcations that result from a border collision between any pair of these equilibria are indicated in \Fig{nsgrid}.

\InsertFig{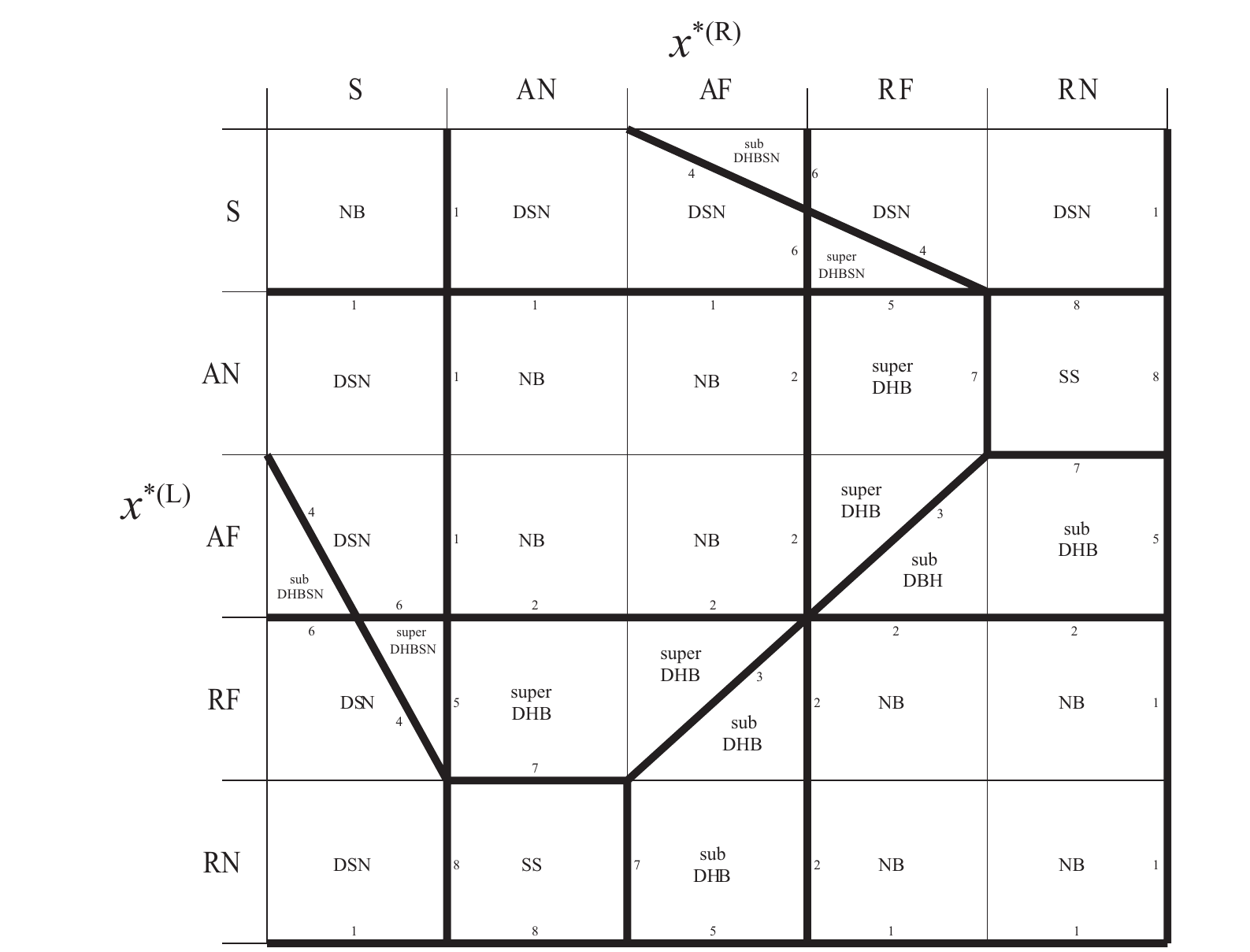}{Codimension-one \db s involving a single \sw~for planar systems,
taken from \cite{Si10}.
Each row [column] represents an equilibrium classification for $f^{(\sL)}$ [$f^{(\sR)}$],
S - saddle;
AN - attracting node;
AF - attracting focus;
RF - repelling focus;
RN - repelling node.
As in \cite{Si10}, names are given to the
five distinct codimension-one phenomena:
NB - no bifurcation;
DSN - discontinuous saddle-node;
DHB - discontinuous Hopf;
DHBSN - discontinuous Hopf-saddle-node;
SS - stability switching bifurcation.
The chosen layout allows for a classification of some codimension-two 
scenarios---these are the thick lines in the figure.
Since smooth parameter variation can produce a change from a saddle
to a repelling node, the left and right sides of the figure
should be considered coincident, as should the top and bottom,
see \cite{Si10} for further details.}{nsgrid}{15cm}

One possible case is that a collision of an equilibrium with a
\sw~yields no bifurcation in the sense that
no invariant sets are created at the collision,
although there is a piecewise-topological change
in the sense of \cite{DiBu08}.
This occurs exactly when the equilibrium
fails to change stability at the bifurcation, and is labeled ``NB" in \Fig{nsgrid}.

When one equilibrium  is a saddle and the other is not,
the two collide and annihilate in a saddle-node-like bifurcation, labeled ``DSN" in the figure. An additional bifurcation may occur if the second equilibrium is a focus: a periodic orbit of
stability opposite to the focus can be created at the \db~\cite{LeNi04,Si10}, this is labeled ``DHBSN".

The \db~also gives rise to a periodic orbit if one equilibrium is attracting,
the other is repelling and at least one is a focus, analogously to a Hopf bifurcation and so labeled ``DHB". 
When both are foci the stability of the generated orbit
is the same as the weaker focus,
where strength is measured by the radial
factor by which the focus
takes trajectories toward or away from the equilibrium upon a rotation of $180^\circ$
\cite{FrPo97,SiMe07}. Thus the criticality of the bifurcation is determined by linear terms in the ODEs, in contrast to the smooth Hopf bifurcation.

The case that $x^{*(\sL)}$ and $x^{*(\sR)}$
are nodes of opposite stability is particularly unusual; 
we refer to the resulting bifurcation as a ``stability-switching" bifurcation,
labeled ``SS" in the figure.
No periodic orbit can be created locally due to the presence 
of the invariant slow manifolds of the nodes.
In the absence of nonlinear terms it is useful to
add a ``circle at infinity'' to the phase plane, an example is shown in \Fig{ppRNAN}.
The resulting \db~is then seen to be analogous to a heteroclinic bifurcation
in that at the singularity there exists a continuous loop
between the equilibrium on the \sw~and two saddles at infinity.
This connection is broken as soon as the equilibrium
of the \pwl~system leaves the \sw.
The presence of nonlinear terms may cause a periodic orbit
to be created at the bifurcation;
however, this orbit will not shrink to the equilibrium at the singularity.
This \db~is seen in, for instance,
a \pwl~version of the FitzHugh-Nagumo model \cite{ArOk97th}.

\InsertFig{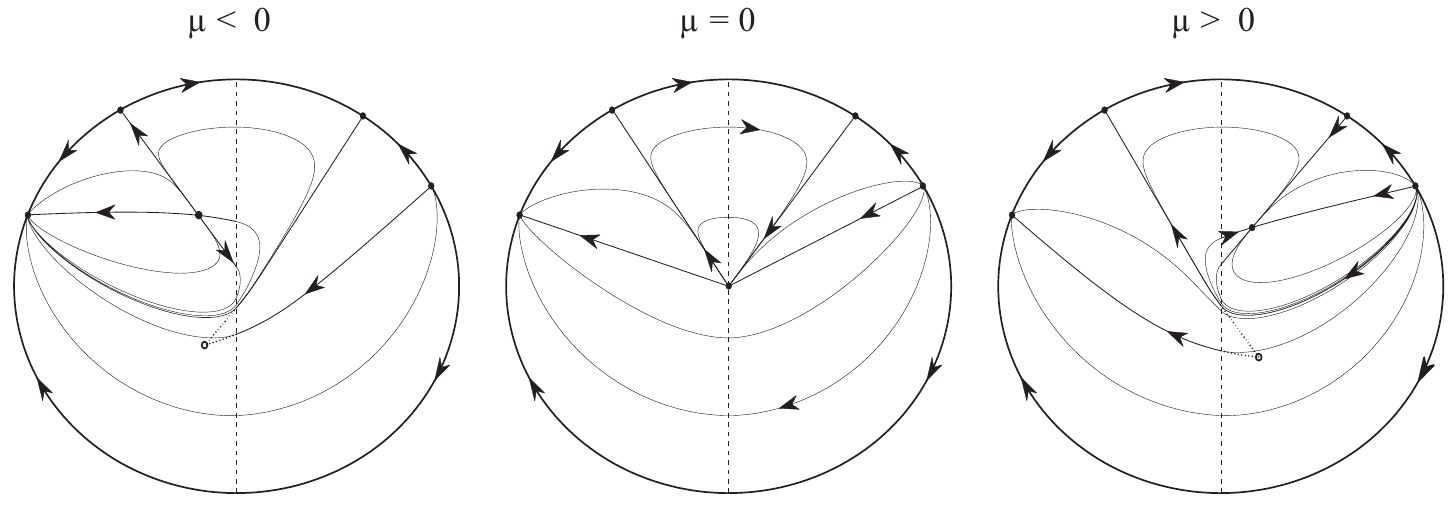}{
Phase portraits including a ``circle at infinity''
for a planar, \pwl, continuous system undergoing
a ``stability-switching'' bifurcation.
The equilibria, $x^{*(L)}$ and $x^{*(R)}$, \Eq{equil},
are nodes of opposite stability
and indicated by hollow circles when virtual.
When $\mu > 0$, the basin of attraction of $x^{*(R)}$
is bounded by a heteroclinic connection between
equilibria at infinity.
The trajectory corresponding to the slow
eigenvector of $x^{*(L)}$ coincides with this connection
for $x<0$; for $x>0$ this trajectory is virtual
and shown by a dotted line.}{ppRNAN}{15cm}

A variety of codimension-two, \db s have been analyzed.
For example, the solid lines in \Fig{nsgrid} 
correspond to codimension-two \db s for planar systems;
their unfoldings are described in \cite{Si10}.
Codimension-one \db s occur along a curve in a
two-parameter bifurcation diagram; an example taken 
from \cite{SiKo09} is shown in \Fig{bsmag2}.
Curves of classical codimension-one, local bifurcations
(such as saddle-node and Hopf bifurcations) involving
an equilibrium of the \db~may have an endpoint at the \db~curve,
like in \Fig{bsmag2}, since beyond this curve the
classical bifurcation is virtual.
Roughly speaking, as one moves along a curve of
saddle-node or Hopf bifurcations, the point in phase space
at which the bifurcation occurs varies linearly.
Equilibria created in a saddle-node bifurcation
move apart from one another by an amount
proportional to the square-root of the parameter change
and for this reason the saddle-node bifurcation curve is tangent
to the \db~curve at its endpoint \cite{SiKo09}.
Similarly the amplitude of the periodic orbit
created in a Hopf bifurcation grows proportional to the 
square-root of the parameter change, hence this Hopf cycle
undergoes grazing along a curve tangent to the Hopf bifurcation
locus at the codimension-two endpoint \cite{SiKo09}.
For planar systems, at the \db~on one side of the codimension-two point
a periodic orbit is created with stability determined by the linear terms
of an appropriate \pwl~expansion.
If this periodic orbit is of opposite stability to the Hopf cycle,
then the two periodic solutions collide in a saddle-node bifurcation
along a curve extremely close to the grazing locus;
specifically, the series expansions of the two curves
agree up to sixth order \cite{SiMe08}.
A rigorous analysis of this bifurcation in higher dimensions,
as well as a multitude of other codimension-two scenarios,
such as the simultaneous occurrence of a \db~with
either a pitchfork or a transcritical bifurcation, remains to be done.

\InsertFig{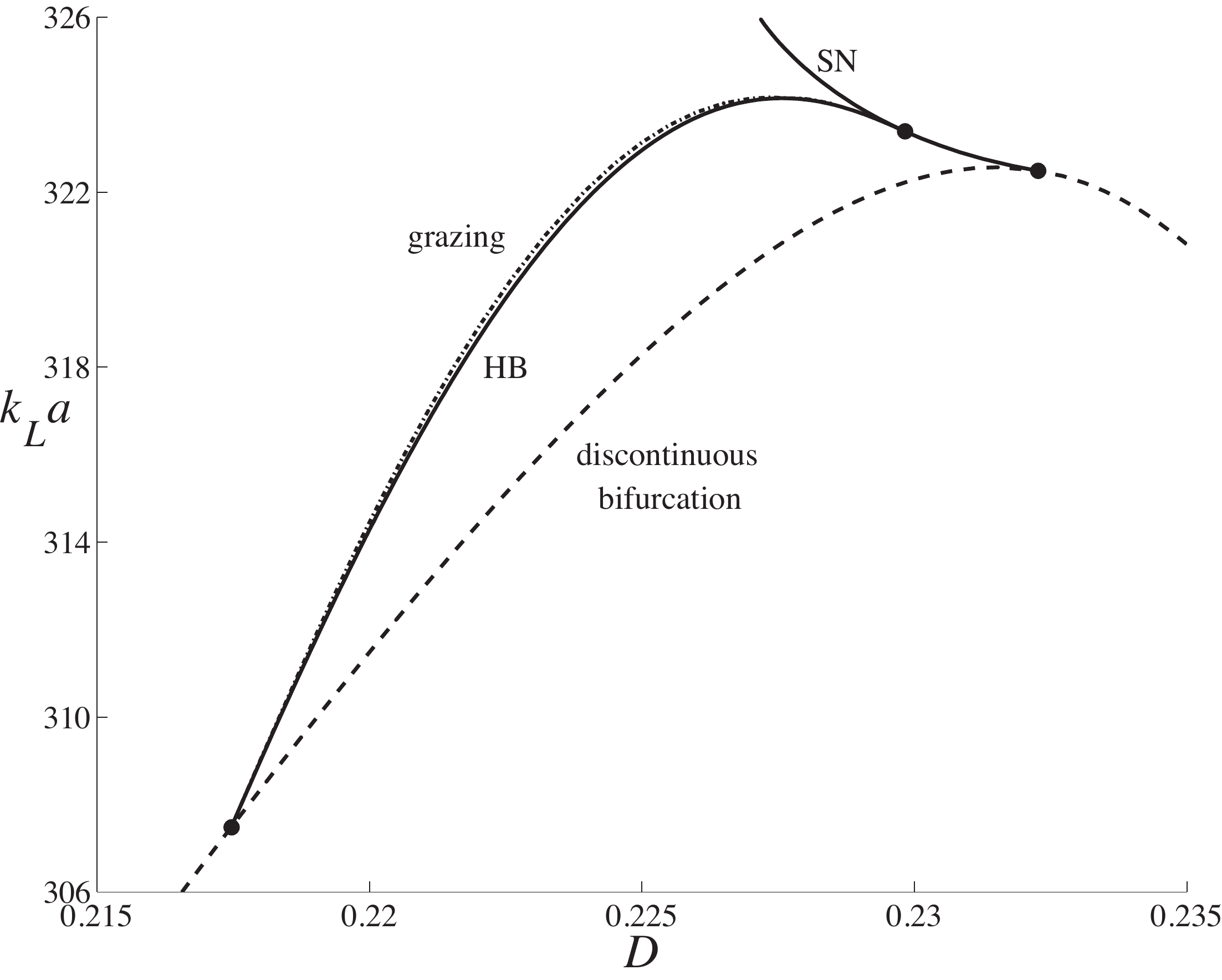}{A two-parameter bifurcation diagram
of a PWSC model of yeast growth, taken from \cite{SiKo09}.
Loci of classical saddle-node (SN) and Hopf (HB) bifurcations emanate
from a locus of \db s.
The Hopf cycle grazes a \sw~along the dash-dot curve
near the Hopf locus.}{bsmag2}{7cm}

\section{Discrete-Time Systems}\label{sec:discreteTime}

As for continuous-time systems,
the stability of fixed points of discrete-time,
PWSC systems is typically not straightforward to determine.
For example the stability of a fixed point at the origin of a planar, \pwl~map may be treated
by reducing it to a circle map with a ``dilation ratio'' assigned to each point 
describing the factor by which
the map takes the point towards or away from the origin \cite{DoKi08}. 
The distribution of the dilation ratios over the natural measure of the circle map
can reveal the stability of the origin.
Of particular interest of late has been the scenario
referred to as a {\em dangerous bifurcation} \cite{HaAb04,GaBa05,Do07}.
This corresponds to the case that the fixed point is stable on
both sides of the \bcb~but is not stable at the bifurcation
and may be analyzed by considering attractors at infinity \cite{GaAv09}.

In one-dimension, the map \Eq{pws} may be written as
\begin{equation}
	x' = \left\{ \begin{array}{lc}
			\mu + a_L x \;, & x \le 0 \\
			\mu + a_R x \;, & x \ge 0
		\end{array} \right. \;,
\label{eq:1dmap}
\end{equation}
and has dynamics that are summarized by \Fig{bs1d},
though it should be noted that unstable solutions are not indicated
in this figure.
The map \Eq{1dmap} can exhibit
discontinuous analogs of saddle-node and Hopf bifurcations for smooth maps,
the primary difference being that
invariant sets separate linearly.
Furthermore, \Eq{1dmap} can have a stable $n$-cycle
(periodic solution of period $n$) for any $n$,
or a chaotic attractor.
For further details see \cite{NuYo95,DiFe99,BaKa00,DiBu08}.

\InsertFig{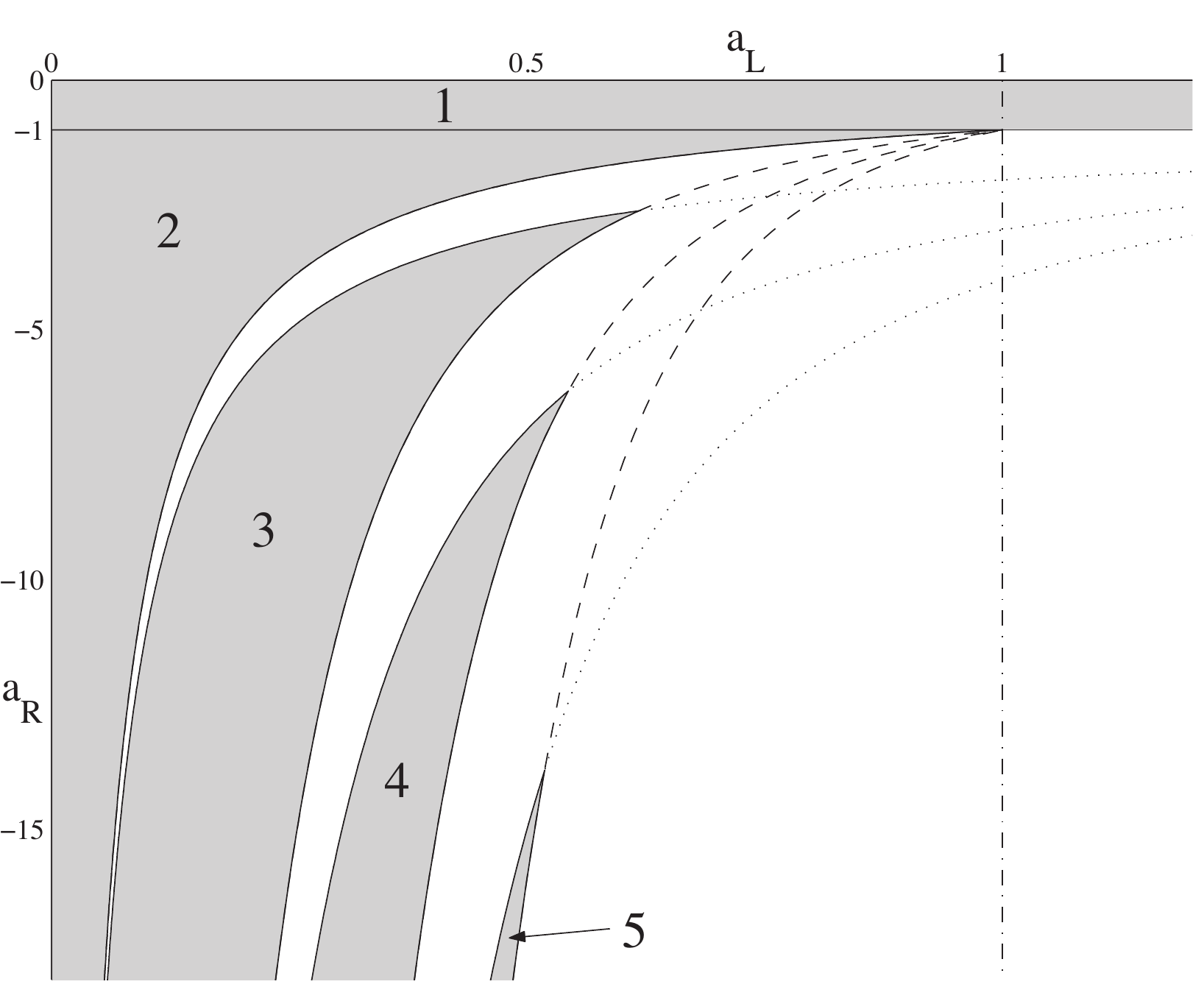}{A two-parameter bifurcation diagram of the 
one-dimensional, \pwl, continuous map \Eq{1dmap}, for $\mu > 0$, taken from \cite{Si10}.
Within each shaded region there exists a periodic solution
of the indicated period.
For $a_L < 1$, there exists a chaotic attractor within the white region.}{bs1d}{8cm}

Dynamics of \pws~maps are often easier to describe
through the use of symbolics \cite{Be00,ChJu04,SuAg05,ZhMo06}.
Since here we are considering only dynamics near a single \sw,
we consider sequences, $\cS$, comprised of only two letters,
$\sL$ and $\sR$.
Any orbit, $\{ x_i \}$, of \Eq{pws} with \Eq{linearDef},
may be assigned a \sew:
\begin{equation}
	\cS_i = \left\{ \begin{array}{lc}
			\sL \;, & {\rm ~if~} s_i < 0 \\
			\sR \;, & {\rm ~if~} s_i > 0
		\end{array} \right. \;,
\end{equation}
where $s_i = e_1^{\sf T} x_i$ and no restriction is placed on $\cS_i$ if $s_i = 0$.

For the piecewise-linear case, $\{ x_0,x_1,\ldots x_{n-1} \}$ is an $n$-cycle if
\[
	x_0 = M_\cS x_0 + P_\cS b \mu \;,
\]
where
\beq{linearSys}
\begin{split}
	M_\cS &= A_{\cS_{n-1}} \ldots A_{\cS_0} \;, \\
	P_\cS &= I + A_{\cS_{n-1}} + A_{\cS_{n-1}} A_{\cS_{n-2}} + \cdots +
	A_{\cS_{n-1}} \ldots A_{\cS_1} \;,
\end{split}
\eeq
Following \cite{SiMe08b,SiMe09,SiMe10},
if $1$ is not an eigenvalue of the matrix $M_\cS$, then
\[
	x_0 = (I-M_\cS)^{-1} P_\cS b \mu \;,
\]
and
\[
	s_0 = \frac{\det(P_\cS)}{\det(I-M_\cS)} \varrho^{\sf T} b \mu \;.
\]
where $\varrho^{\sf T} = e_1^{\sf T} \adj(I-A_\sL(\mu))$ 
(as implied by the discussion of \Eq{rhodefine}, which is stated for
continuous-time systems).

These algebraic methods can be used to find admissible periodic solutions, and hence
to find regions in parameter space
where periodic solutions exist for some fixed $\mu$. 
As in the smooth case, these regions are known as resonance or Arnold tongues,
see \Fig{resrLwR}.
Boundaries of resonance tongues correspond to
the corresponding $n$-cycle losing stability or
undergoing a border-collision.
In the latter case one point of the $n$-cycle typically collides with the \sw.
Remarkably a \pwl~expansion of the appropriate $n^{\rm th}$-iterate
of the map will also have the form \Eq{pwl}, i.e.,~exactly the same form
as the underlying map \cite{DiFe99,Si10}.
Hence \Eq{pwl} may also be used to analyze \bcb s of $n$-cycles.
The stable and unstable manifolds of periodic saddles may undergo
homoclinic tangencies leading to homoclinic tangles
and chaos \cite{ZhMo06b}.

\InsertFig{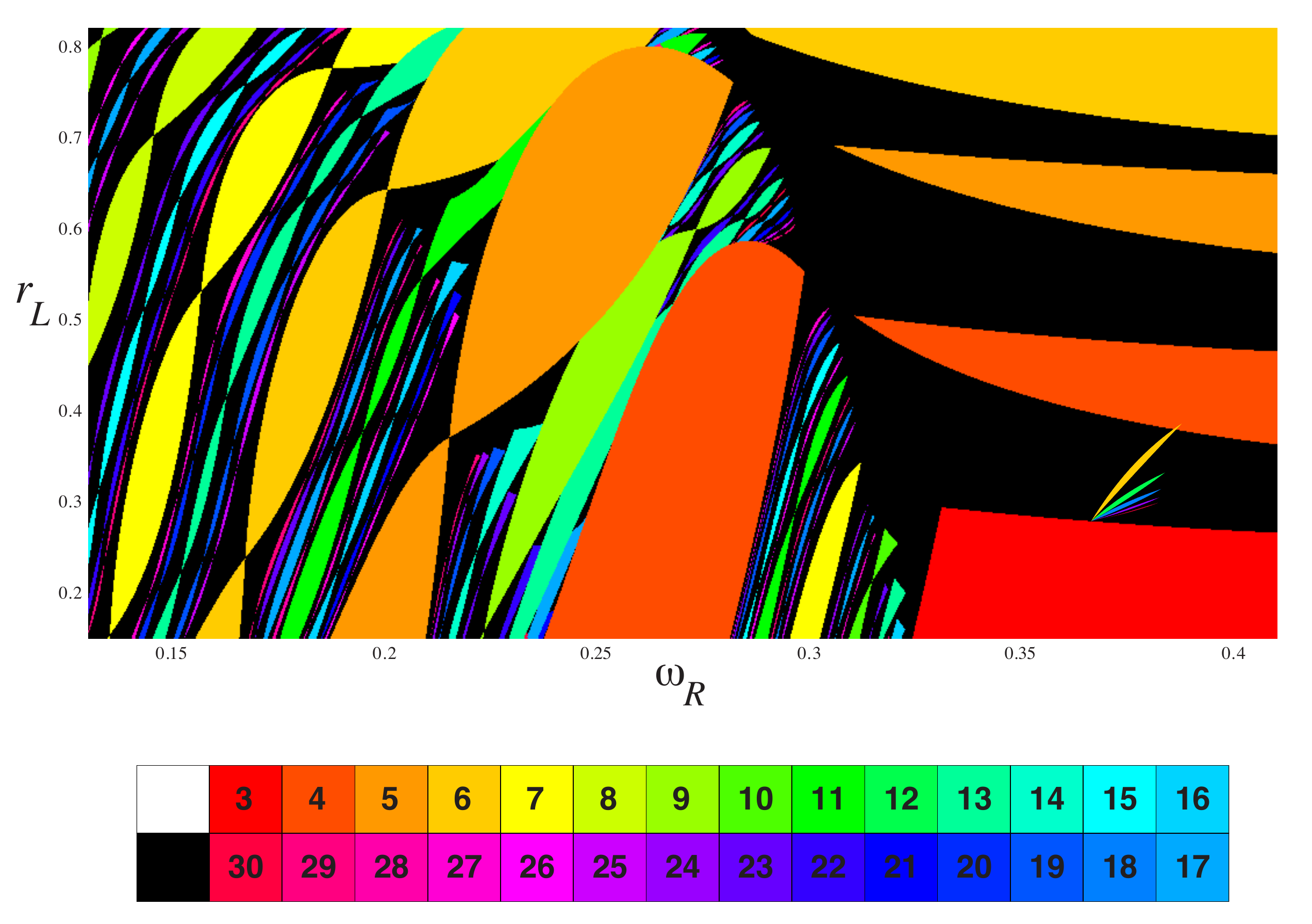}{
Resonance tongues of \Eq{pwl} with \Eq{companion} for $\mu > 0$
where the multipliers of $C_\sL(0)$ and $C_\sR(0)$ are
$r_L {\rm e}^{\pm 2 \pi {\rm i} \omega_\sL}$ and
$\frac{1}{\sR} {\rm e}^{\pm 2 \pi {\rm i} \omega_\sR}$, respectively,
and $s_\sR = 0.8$, $\omega_\sL = 0.09$, taken from \cite{Si10}.
Period is indicated by the color bar.
Within the black regions, forward orbits either approach
an orbit of period greater than $30$ or are aperiodic.}{resrLwR}{12cm}

Resonance tongues in a generic two-parameter bifurcation diagram, like \Fig{resrLwR}, for \pwl, continuous maps often display a structure resembling a string of sausages \cite{YaHa87,CaGa96,SuGa04,PuSu06,ZhMo06b,SiMe08b,SzOs09}.
A point at which a tongue has zero width is known as a {\em \shr}.
When the \sew~of the periodic orbit is {\em rotational},
as defined in \cite{SiMe09},
the structure near a shrinking point is well understood.
Rotational symbol sequences are natural as they correspond to rigid translations on an invariant circle. A rigorous unfolding of \shr~bifurcations for rotational periodic orbits can be given under reasonable nondegeneracy conditions. The analysis shows that four distinct border-collision curves emanate from a \shr; each is tangent to one of the others at the \shr, an example is shown in \Fig{rr17}.
For a nonlinear, PWSC map, shrinking points occur only in the limit $\mu \to 0$. As $\mu$ grows, they break apart through the formation of curves of classical saddle-node bifurcations, as shown in \Fig{PWSCShr} \cite{SiMe10}.
The existence of quasiperiodic orbits with irrational
rotation numbers can be inferred from the numerics
as limits of periodic orbits \cite{SiMe08b};
however, these have not been treated rigorously as of yet,
except in some special cases \cite{SzOs08,LaRa05b}.

\InsertFig{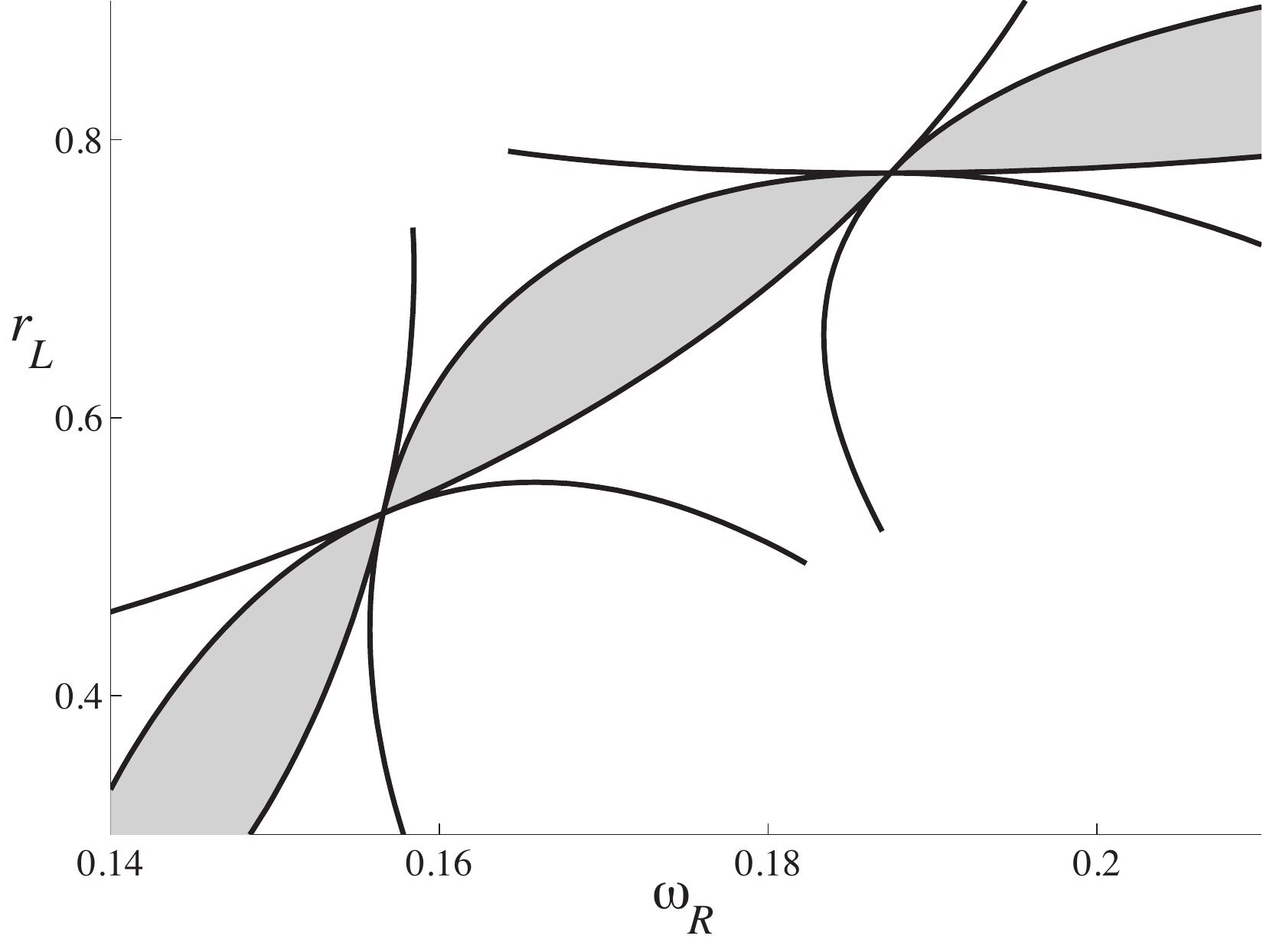}{
A magnification of the \tong~in \Fig{resrLwR}
corresponding to a rotation number of $1/7$.
The boundary of the \tong~is defined
by several curves along which there exists
a $7$-cycle with one point on the \sw.
}{rr17}{8cm}

\InsertFig{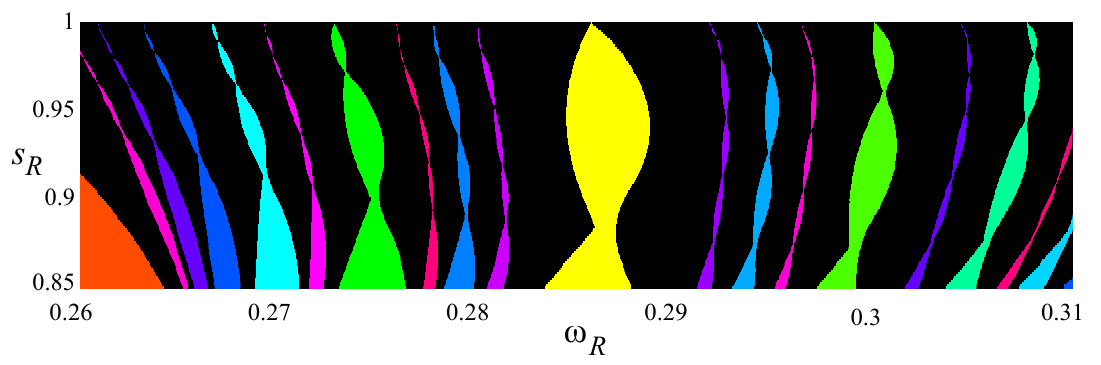}{Resonance tongues for a PWSC map with multipliers and color scheme as in \Fig{resrLwR}, taken from \cite{SiMe10}. Here $r_L = 0.2$, $\omega_L = \omega_R$, $\mu = 0.2$ and the map $f^{(\sL)}$ has a quadratic nonlinearity. The right boundary of each resonance tongue has become $C^1$ through the creation of a curve of saddle-node bifurcations, and the non-terminating \shr s have been destroyed.}{PWSCShr}{5in}

The upper boundary of the period-three tongue (the red region in the lower right corner of \Fig{resrLwR})
corresponds to a loss of stability of a $3$-cycle
when a multiplier becomes $-1$.
At this boundary, a $6$-cycle is created, though it is unstable
and thus not shown in the figure,
for further details see \cite{MaSu98,SuGa06,SiMe08b,Si10}.
Near the point $(\omega_\sR,r_\sL) \approx (0.366,0.278)$ 
small, higher-period \tong s emanate from a special point on the boundary
of the period-three tongue
in a manner that is only partially understood \cite{SiMe08b,Si10}.

If one of $A_{\sL}(0)$ and $A_{\sR}(0)$ for \Eq{pws}
has a multiplier on the unit circle, the \bcb~at $\mu = 0$ is degenerate.
The three codimension-two cases---a multiplier $1$,
a multiplier $-1$, or a complex pair of multipliers
with unit modulus---correspond to the coincidence
of a \bcb~with a saddle-node,
a period-doubling,
or a Neimark-Sacker bifurcation, respectively. 
Several unfoldings of these bifurcations are given in \cite{Si10}.
Recently these three scenarios have been unfolded
by Colombo and Dercole \cite{CoDe10}
without the assumption that the map is continuous,
in fact without any information about the map on one side of the \bcb.
The results mirror those described in \Sec{continuousTime} for the continuous-time case.
With reasonable nondegeneracy assumptions,
loci of saddle-node bifurcations and \bcb s intersect tangentially,
whereas loci of period-doubling and Neimark-Sacker bifurcations
intersect the \bcb~loci transversely.
Since the newly created solutions along period-doubling and Neimark-Sacker bifurcation loci
grow in size as the square-root of a parameter, they
consequently undergo border-collision along a tangent curve
as sketched \Fig{BCBcodim2}.
The nature of the bifurcation at this additional border-collision
depends on the particular map.
For example, it can be shown for the one-dimensional case that an attracting chaotic solution may be created along the period-doubling, border-collision curve \cite{SiMe09b}.

\InsertFig{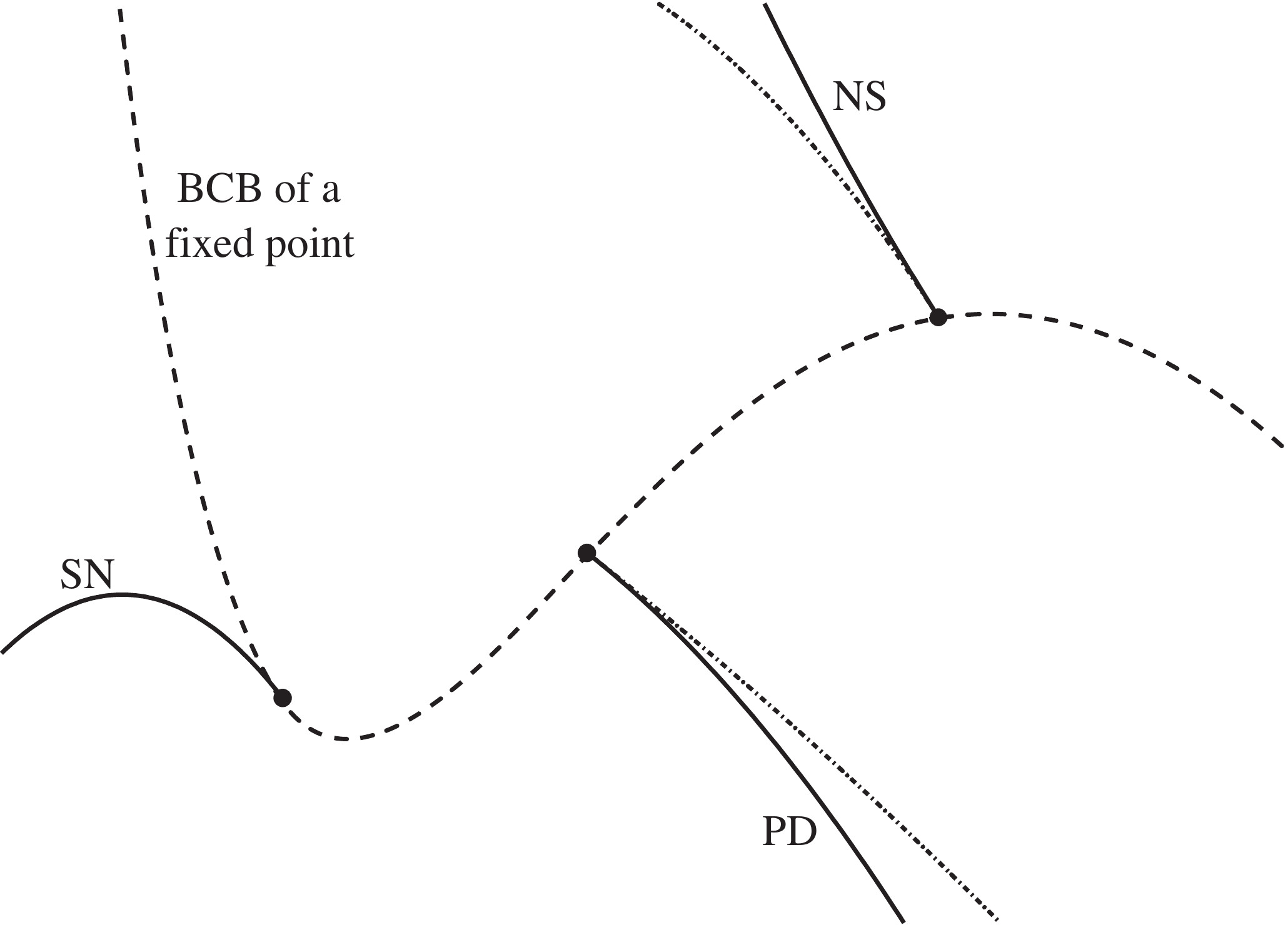}{A schematic two-parameter bifurcation diagram
for a PWSC map.
The long dashed curve corresponds to a \bcb~of a fixed point.
The solid curves correspond to loci of classical saddle-node (SN),
period-doubling (PD) and Neimark-Sacker (NS) bifurcations.
The dash-dot curves correspond to the collision
of the period-doubled solution and the invariant circle
with the \sw. The tangency or transversality of these curves follows from the general unfolding in \cite{Si10, CoDe10}. }{BCBcodim2}{7cm}

\section{Discussion}\label{sec:discussion}

Dynamics related to the interaction of an equilibrium
in a PWSC flow, or a fixed point in a PWSC map,
with a smooth \sw~are
determined by a piecewise series expansion
\Eq{pws} with \Eq{linearDef},
and, except in special cases, only the linear terms
of this expansion are important.
Consequently invariant sets created at a
discontinuous or \bcb~grow linearly,
to lowest order, with respect to parameter change.
Arguably this is the most important feature
that distinguishes discontinuity induced bifurcations with
bifurcations of smooth systems.
For example this difference has recently been found in a cardiac model
\cite{ZhSc07,ZhSc08} suggesting that a \bcb~and not
a classical period-doubling bifurcation
generates the observed period-doubled solution.

Discontinuous bifurcations in one and two-dimensional systems
are well understood, as are simple \bcb s.
Often the local dynamics is determined by global properties
of the appropriate \pwl~approximation. A number of codimension-two bifurcations
that correspond to the  coincidence of
a discontinuous or \bcb~with one of several classical bifurcations
have also been unfolded.
However, several codimension-two scenarios remain to be fully explored,
such as the coincidence of border-collision and Neimark-Sacker bifurcations.

A discontinuity induced bifurcation in a high-dimensional system
could be extremely complicated --- or it could behave like
a bifurcation in low-dimensional system.
In the latter case, a more complete understanding
of the bifurcation could result from
some sort of dimension reduction; however, as of yet no widely applicable theory for dimension reduction
exists for PWSC systems.

Resonance tongues in \pwl, continuous maps
often have points of zero width (\shr s)
that are located at the intersection of several border-collision curves. While much of
the structure of \shr~bifurcations has been explored, there are still a number of open questions. For example, the theory only applies to \shr s in rotational symbol sequences; are there similar bifurcations for non-rotational orbits?
What differences arise when the lowest order nonlinear terms are not quadratic,
but rather some fractional power
(as in maps derived from sliding bifurcations in Filippov systems)?
Also the observation in \cite{SiMe08b} that curves of resonance tongue \shr s may bound chaotic behavior remains to be explained.
Finally, at a \shr, the map has an invariant polygon that may continue as an invariant circle.
This circle can be destroyed by various mechanisms
\cite{SuGa06,ZhMo08,ZhSo07}; however, analogies of a number of the rigorous results for the smooth case remain to be obtained for PWSC maps,
perhaps due, in part, to a lack of normal hyperbolicity.

\bibliographystyle{model1-num-names}
\bibliography{PWSC}

\end{document}